\documentstyle[aps,prd,amssymb,eqsecnum,epsf,twocolumn]{revtex}
\newcommand{\be}{\begin{equation}}
\newcommand{\ee}{\end{equation}}
\newcommand{\bea}{\begin{eqnarray}}
\newcommand{\eea}{\end{eqnarray}}
\newcommand{\nn}{\nonumber}

\newcommand{\newc}{\newcommand}
\newc{\ra}{\rightarrow}
\newc{\lra}{\leftrightarrow}
\newc{\beq}{\begin{equation}}
\newc{\eeq}{\end{equation}}
\newc{\ba}{\begin{eqnarray}}
\newc{\ea}{\end{eqnarray}}
\newskip\humongous \humongous=0pt plus 1000pt minus 1000pt

\newif\ifdtup

\relax
\title{Metastability of Spherical Membranes in Supermembrane and Matrix Theory}
\author{M. Axenides, E. G. Floratos and L. Perivolaropoulos}
\address{Institute of Nuclear Physics, \\
National Centre for Scientific Research ``Demokritos N.C.S.R.'',\\
Athens, Greece\\ e--mail: {\tt axenides@mail.demokritos.gr, \
manolis@mail.demokritos.gr, \ leandros@mail.demokritos.gr}}

\date{\today}
\draft

\begin{document}

\maketitle

\begin{abstract}
Motivated by recent work we study rotating ellipsoidal membranes 
in the framework of the light-cone supermembrane theory. We 
investigate stability properties of these classical solutions 
which are important for the quantization of super membranes. We 
find the stability modes for all sectors of small multipole 
deformations. We exhibit an isomorphism of the linearized membrane 
equation with those of the $SU(N)$ matrix model for every value of 
$N$. The boundaries of the linearized stability region are at a 
finite distance and they appear for finite size perturbations. 
\end{abstract}

\pacs{PACS:  }

\section{Introduction}
M theory\cite{TWSD} is considered today as the best candidate for
the unification of the weak and strong coupling sectors of all
known string theories. The most serious attempt up to now to frame 
M-theory together with all of its ingredients is the Matrix 
theory\cite{BFSS}. The conjecture of this theory is that all the 
 freedom of various sectors of the five known string theories can be represented
by appropriate operators of the Matrix theory using duality
properties. Up to now all perturbative checks of this idea
although not straight forward have
 been sucessfull and focus on the nonperturbative sector leads to the
study of classical solutions of Matrix theory in general
backrounds. Recent progress in this direction is the successfull
formulation of Matrix theory in weak gravitational and gauge
backgrounds i.e. dynamics of matrix branes in the backround of
their mutual and external forces\cite{RCM}.

One of the poorly understood elements of M-Theory is the eleven
dimensional classical supermembrane sector. Progress in this
direction is important both for the understanding of the strongly
coupled string theories as well as for the quantization of the
supermembrane. Recent interest for the classical solutions of the
Matrix theory representing $D_0$-branes attached to spherical
membranes is explained as a first step to a formulation of Matrix
theory in weak external gravitational and gauge backrounds
\cite{KT}. Particular solutions of the classical matrix equations
representing rotating ellipsoidal configurations of N $D_0$ branes
attached to a membrane which exhibit stability properties have
been proposed and their semiclassical spectrum has been studied
\cite{HS}.

In this work we study in detail the stability properties of
rotating spherical membrane which are solutions of the bosonic
part of the supermembrane equations restricted to six spatial
dimensions. They are motivated by the recently found matrix model
solutions which represent N $D_0$-branes pinned on the surface of
 a rotating ellipsoidal membrane\cite{DKPS}.
We find stability for all modes of small multipole deformations 
and we determine explicitly the spectrum and the eigenmodes. There 
is an interesting isomorphism with the full stability analysis of 
the Matrix solution which demonstrates that classical membrane 
excitations can be analyzed in distinct 
 sectors of $D_0$-branes and provide approximations for the quantum mechanical
study of the spherical membrane.% In addition a particular
%"fissioning" mechanism of rotating spherical membrane is suggested
%by the details of the classical instabilities which lead to a
%proposal for membrane vertex operators which can be usefull in
%first quantized interacting supermembrane theory\cite{Nicolai}.

\section { Matrix Models versus Supermembrane}
It is a well known fact that the matrix model was one of the first
ideas to study the bosonic membrane in the light cone frame in the
approximation of finite number of oscillation modes. The elegant
observation of \cite{WHN} is that $SU(N)$ Yang-Mills mechanics is
a consistent mode truncation of the membrane excitations in the
light cone frame and moreover in the late eighties when the
supermembrane Lagrangian\cite{BST}  was written down it became
clear that the dimensional reduction of the ten dimensional $N=1$
SUSY Yang-Mills theory to $d=1$ is the correct supersymmetric
extension of the above truncation.\cite{WHN} The light cone
infinite dimensional area preserving symmetry of the supermembrane 
Hamiltonian \cite{WHN,BST}is truncated by $SU(N)$ Yang-Mills 
symmetry.  This can be represented as the algebra of the 
corresponding discrete and finite Heisenberg group of a 
discretized membrane considered as a two dimensional discrete 
phase space\cite{F}. The large $N$ limit connecting $SU(N)$ 
Yang-Mills to the membrane Hamiltonian, i.e. commutators with 
Poisson brackets became clear as an analogy of the passage from 
quantum to classical mechanics $\hbar={{2\pi}\over N}\rightarrow 
0$, as $N\rightarrow \infty$\cite{FIT}. With regard to the study 
of the quantum theory of the membranes it is necessary to 
understand better this limit both from the point of view of matrix 
models in general but also from the approximation point of view of 
the measure for the quantum configuration \cite {N,FL}.

We now make a quick review of the formalism for the bosonic sector
of the theory relevant to the present work. After fixing the gauge
and using reparametrization invariance of the Nambu-Gotto
Lagrangian we find that the eqs of motion for the $9$ bosonic
coordinates $X_{i}(t,\sigma_{1},\sigma_{2})$, $i=1,2,... ,9$ in 
the light cone frame are: \beq \label{E1} 
\ddot{X}_{i}\;=\;\{X_{k},\{X_{k},X_{i}\}\} 
 \eeq
  where the Poisson bracket of two functions , $f$ and $g$ on $S^{2}$
  is defined as
 \beq \label{E2}
\{f,g\}\;=\; \frac{\partial {f}}{\partial{\cos \theta}}
\frac{\partial {g}}{\partial{ \phi}} \;-\;\frac{\partial
{f}}{\partial{\phi}}\frac{\partial {g}}{\partial{\cos\theta}} \eeq
and the remaining area preserving symmetry generated by the
constraint
 \beq\label{E3}
\left\{ X_{i}, {\dot X}_{i}\right\}= 0  \eeq In the matrix model
the above coordinates are replaced by $N \times N$ Hermitian and
traceless matrices and the corresponding equations of motion and
constraint are found by exchanging Poisson brackets with
commutators.

The first connection between the $SU(N)$ Susy Yang-Mills
truncation of the supermembrane with the recent nonperturbative
studies of string theories was discovered by Witten \cite{W96}
representing the Yang-Mills mechanics as a low energy effective
theory of bound states of $N$ $D_{0}$ branes. The $D_{0}$ branes
carry RR charge. Now it is understood how to couple the $SU(N)$
matrix model with weak background fields either directly using
supergravity arguments or truncating supermemmbrane Lagrangians in
weak background fields\cite{RCM}. There is an expectation that
taking appropriate limits of $N\rightarrow\infty$ for special
bound states of $N$ $D_{0}$ branes one could recover the
supermembrane or its magnetic dual, the super-five brane\cite{PT}.
On the other hand the study of classical solutions of
supermembranes or matrix model could provide a nonperturbative
information for their dynamics even in the quantum regime. In the
next section we turn our attention to the analysis of the
stability properties of specific classical solutions which are
spherical rotating membranes. Recent work in the matrix model 
presented such a time dependent solution representing a bound 
system of $N$ $D_{0}/D_{2}$ branes.

\section{Spherical Membranes as Matrices}
Since we are going to study spherical membranes and their matrix
analogs, we begin by reviewing the salient features of the Lie
algebra $sDiff(S^{2})$ of area preserving diffeomorphisms of the
sphere $S^{2}$ considered as a two dimensional differentiable
manifold. We first introduce the canonical or Darboux coordinates
on the sphere $ \sigma_{1}=\phi$ , $\sigma_{2}=\cos\theta$. In
these coordinates the Poisson bracket on the sphere is defined as

\beq\label{E4}
 \{f,g\}\; = \;
 \frac{\partial f}{\partial
\sigma_1} \frac{\partial g}{\partial{\sigma_{2}}} \;-\;
\frac{\partial f}{\partial{\sigma_{2}}}\frac{\partial g}{\partial
{\sigma_{1}}}
 \eeq
where $f,g \in \cal C^{\infty}(S^{2})$.  The spherical harmonics
$Y_{l,m}(\theta,\phi)$ for $m =-l,\ldots
 l$ and $l=0,1,\ldots,\infty$ give rise to a complete system of
 generators for $sDiff(S^{2})$:

\beq\label{E5}
 L_{l,m}\; = \;\frac{\partial Y_{l,m}}{\partial
{\cos\theta}}\frac{\partial}{\partial{\phi}} \;-\; \frac{\partial
Y_{l,m}}{\partial{\phi}}\frac{\partial}{\partial {\cos \theta}}
 \eeq

%The generators of symplectic transformations of a surface locally
%have the form \beq\label{E6}
% L_{f}\; =\; \frac{\partial f}{\partial
%\sigma_2}\frac{\partial{} }{\partial\sigma_{1}}\;-\;\frac{\partial
%f}{\partial\sigma_{1}}\frac{\partial{}}{\partial \sigma_{2}} \eeq
%and they satisfy the algebra \beq\label{E7}
% [L_{f}, L_{g}] \;=\;
%L_{{f,g}} \eeq
They satisfy the algebra
 \beq \label{E8}
[L_{l,m},L_{l^{'},m^{'}}]\;=\;
f^{l^{"}m^{"}}_{lm,l^{'}m^{'}}\;\;L_{l^{"}m^{"}} \eeq where the
structure constants  $ f ^{l^{"}m^{"}}_{lm,l^{'}m^{'}}$ are
defined by expanding the Poisson brackets on the basis $Y_{l,m}$
as follows:

\beq\label{E9}
 \{Y_{l,m}\;,\;Y_{l^{'}m^{'}}\}\;=\;f
^{l^{"}m^{"}}_{lm,l^{'}m^{'}}\;Y_{l^{"}m^{"}} \eeq
 These structure constants have been calculated explicitly by Hoppe in Ref.
\cite{WHN}. In order to get a feeling of the geometrical meaning
of these generators , we note that for $l=1, m=0,\pm 1$, these are
the usual angular momentum generators (up to normalization),
$L_{z},L_{\pm}$. Also \beq\label{E10}
L_{l,m}\;Y_{l^{'},m^{'}}\;=\;- \{ Y_{l,m} , Y_{l^{'},m^{'}}\}\;=\;
- f^{l^{"}m^{"}}_{lm,l^{'}m^{'}}\; Y_{l^{"},m^{"}} \eeq so that
\beq\label{E11}
 [L_{1,\pm1}, L_{l,m}]\;\cong\; [l(l+1) -
m(m+1)]^{1/2} L_{l,m\pm1} \eeq \beq\label{E12}
[L_{1,0},L_{l,m}]\;\cong\;m\;L_{l,m} \eeq For general l,m ,
$L_{l,m}$ produce multipole deformations of the spherical
membrane.
 The above two eqs
imply that the infinite set of generators $L_{l,m}$ is reduced to
an infinite sum of irreducible sets of operators of definite
angular momentum l, \beq\label{E13} L_{l,-l},L_{l,-l+1},\ldots,
L_{l,l},\;\;\;\;\;l=1,2,\ldots \eeq with respect to the $SU(2)$
Lie algebra, $L_{1,0},L_{1,\pm 1}$ of solid rotations on the
sphere. We now describe the approximation of $sDiff(S^{2})$ by
$SU(N)$. We choose an appropriate basis of $SU(N)$ describing
matrix spherical harmonics \cite{WHN}. The spherical harmonics
$Y_{l,m}(\theta,\phi)$ are harmonic homogeneous polynomials of
degree $l$ in the three euclidean coordinates $x_{1},x_{2},x_{3}$
of points on $S^2$  where: \beq\label{E14}
 x_{1}=\cos\phi
\sin\theta , x_{2}=\sin\phi\sin\theta, x_{3}= \cos\theta \eeq and

\bea \label{E15}Y_{l,m}(\theta,\phi) &=&  \sum_{ i_{k}=1,2,3}
\;\alpha^{(m)}_ {i_{1},\ldots,i_{l}}\;x_{i_{1}},\ldots,x_{i_{l}},
\nn
\\k&=&1,\dots,l \;\; k=1,\dots,l \eea
 where $ \alpha^{(m)}_{i_{1},\ldots,i_{l}}$ is a symmetric and
traceless tensor. For fixed $l$ there are $2l+1$ linearly
independent tensors $ \alpha^{(m)}_{i_{1},\ldots,i_{l}}$,
$m=-l,\ldots,l$ \cite{Ham}.

Let $J_{1},J_{2},J_{3} $ be $N \times N$ hermitian matrices which
form an N-dimensional irreducible representation  of the Lie
algebra $SU(2)$, \beq\label{E16}
  [J_{i}, J_{j}]\;=\; i \epsilon _{ijk} J_{k}
\eeq  Hoppe in  Ref \cite {WHN} has  shown that the matrix
polynomials \beq\label{E17}
 \hat{Y}^{(N)}_{l,m}\;=\; \sum_{i_{k}=1,2,3}
\alpha^{(N)}_{i_{1},\ldots,i_{l}}\; J_{i_{1}}\ldots J_{i_{l}} \eeq
for $l=1,\ldots,N-1,\; m=-l,\ldots,l$ can be used to construct a
basis of $N^{2} -1 $ matrices  for the fundamental representation
of $SU(N)$ with corresponding structure constants $f^{(N)}$:
\beq\label{E18}
 [\hat{Y}^{(N)}_{l,m}, \hat{Y}^{(N^{'})}_{l^{'},m^{'}}]\;\;=\;\;i f^{(N)l^{"}m^{"}}_{lm,l^{'}m^{'}}
 \hat{Y}^{(N)}_{l,m}. \eeq

There is a normalization of the generators $ \hat{Y}^{(N)}_{l,m}$
such that the limit \beq\label{E19}
Nf^{(N)l^{"}m^{"}}_{lm,l^{'}m^{'}}\stackrel{N\rightarrow\infty}{\longrightarrow}
 f^{l^{"}m^{"}}_{lm,l^{'}m^{'}}
 \eeq
exists and coincides with the structure constants
 as defined before in eq:(\ref{E9}). After these
preliminaries we proceed to establish the relation of the infinite
dimensional algebra  eq(\ref{E8}) ,$sDiff(S^{2})$, to the $SU(N)$
algebra as $N\rightarrow\infty$ , by an argument which avoids the
explicit computation of ref \cite{WHN} for the structure constants
$f^{(N)}$ and $f$\cite{FIT}. If we rescale the generators of
$SU(2)$ by $1/N$ \beq\label{E20}
 J_{i}\rightarrow T_{i}\;=\; (1/N) J_{i}
 \eeq

they satisfy the algebra \beq\label{E21}
 [T_{i}, T_{j}]\;=\;
(\imath/N)\epsilon_{ijk}T_{k} \eeq and the Casimir element
\beq\label{E22} T^{2}\;=\; T^{2}_{1} + T^{2}_{2} +
T^{2}_{3}\;\simeq\; 1 + 1/N \eeq has a finite limit for
$N\rightarrow\infty$. Under the norm \beq \label{E23}
 |x|^{2} \equiv
Tr(x^2), \eeq for $x\in SU(2)$, the generators $T_{i}, i=1,2,3$ 
have definite limits as $N\rightarrow\infty$ which are three 
objects $x_{1},x_{2},x_{3}$ which commute and are constrained by 
eq.(\ref{E22}) according to \beq\label{E24} 
 x^{2}_{1} +
x^{2}_{2} + x^{2}_{3} = 1 \eeq
 If we consider two polynomial
functions of three commuting variables $f(x_{1},x_{2},x_{3})$ and
$g(x_{1},x_{2},x_{3})$ the corresponding matrix polynomials
$f(T_{1},T_{2},T_{3}), g(T_{1},T_{2},T_{3})$ have commutation
relations for large N which follow from eq(\ref{E21}): \beq
\label{E25} lim_{N\rightarrow\infty}( N/i)[f,g]\;=\;
\epsilon^{ijk}x^{j}\frac{\partial f}{\partial x^{j}}\frac{\partial
g}{\partial x^{k}} \eeq

This is similar to the passage from quantum mechanics to classical
mechanics. This can be generalized to all semisimple Lie groups.
 If we
parametrize $x_{i}$ by polar coordinates (see eq(\ref{E14})) we 
see that the right hand side of the previous equation is nothing 
else but the Poisson brackets. Consider now the basis 
$T^{(N)}_{l,m}$ of $SU(N)$ obtained by replacing in (\ref{E17}) 
the matrices $J_{i}$ by the rescaled ones $T_{i}$. Then according 
to (\ref{E20}) we obtain \beq\label{E26} 
\lim_{N\rightarrow\infty}\frac{N}{\imath}[T^{(N)}_{l,m}, 
T^{(N)}_{l^{'},m^{'}}]\;=\; \{ Y_{l,m},Y_{l^{'},m^{'}}\} \eeq 
 If we
replace the left hand side with eq (\ref{E18}) we obtain eq
(\ref{E19}). From the above
 discussion it is obvious that the membrane equations of motion
 and constraint are the semiclassical limit $N\rightarrow\infty$
 of the corresponding matrix equations.
 Going from the membranes to the matrix model is analogous to the
 correspondence of classical with quantum mechanics. The various
 observables of the classical membrane correspond to $N\times N$
 matrices but there are ordering ambiguities.

 Below we present an explicit construction of a
 completely symmetrized basis of observables
 in the matrix model which corresponds to the basis of spherical
 harmonics as was pointed out by eq.(\ref{E15}). This method was first developed by
 Schwinger\cite{Sch}.

 Let ${\bf \alpha}$
 a complex null three dimensional vector i.e. $\alpha^{2} =\alpha
 \cdot\alpha=0$ parametrized by two complex numbers $z_{+},z_{-}$
 \beq
 \alpha_{1}\;=\;-z^{2}_{+} + z^{2}_{-},\;\;\; \alpha_{2} =
 -\imath(z^{2}_{+} + z^{2}_{-}), \alpha_{3} = 2 z_{+}z_{-}
 \eeq

If {\bf r } as the position vector then $( a\cdot r)^k$ is a
spherical harmonic of order k given by :
 \beq\label{E27}
  \frac{(a\cdot r)^{k}}{2^{k}k!}\;=\; \left[\frac{4 \pi}{2k +1}
  \right]^{1/2} \sum_{m}\Phi_{jm}(z) Y_{jm}( r) \eeq
where  \beq \Phi_{jm}(z)\;=\; \frac{ z^{j+m}_{+} z^{j-m}_{-}} 
{[(j+m)!(j-m)!]^{1/2}} \eeq and also $Y_{jm}({\bf r})$ usually 
designates a spherical harmonic. Here it includes a factor 
$r^{k}$. Accordingly 
 \beq\label{E28}
  \frac{(a\cdot J)^{k}}{2^{k}k!}\;=\; \left[\frac{4 \pi}{2k +1}
  \right]^{1/2} \sum_{m}\Phi_{jm}(z)\hat{Y}_{jm}( J) \eeq
 in which $\hat{Y}_{jm}( J)$ differs from the analogous $Y_{jm}
(r)$ only in that the order of factors is significant.

   These
 operators have the following properties :
\beq\label{E29}
 \hat{Y}_{jm}({\bf J})^{\dagger}\;=\; (-1)^{m}\hat{Y}_{j -m}({\bf J}). \eeq

If J belongs to  the spin j representation they also satisfy an 
 orthogonality and tracelessness property given by eqs. \beq \label{E31} 
 \frac{1}{2j+1} tr \hat{Y}_{j_{1}m_{1}} ({\bf J})^{\dagger}
\hat{Y}_{j_{2}m_{2}}({\bf J})\;=\; \frac{1}{4\pi} [{j
(j+1)}]^{j_{1}} \delta_{j_{1}j_{2}} \delta_{m_{1}m_{2}} \eeq

 \beq\label{E32}
\frac{1}{2j+1} tr \hat{Y}_{jm}( J)\;=\; \delta_{j0} \eeq

The $ N\times N$ matrices $\hat{Y}_{lm}$ are nothing else but the
standard spherical tensor operators of Quantum Mechanics\cite{Ba}

\section{Stability}

The equation of motion for the supermembrane in six dimensions may
be written as

\begin{equation}
  \label{eq:motion}
  \ddot{X}_i= \left\{ X_j,\left\{X_j, X_i \right\}\right\}
\end{equation}
where summation is implied in the j indices and $\left\{ \right\}$
stands for the Poisson bracket with respect to the angular
coordinates $\theta, \; \phi$. The Gauss constraint that also
needs to be satisfied is
\begin{equation}
  \label{eq:gauss}
 \left\{\dot{X}_i, X_i \right\}=0
\end{equation}
where $i,j=1,2..6$. We now define $Y_i\equiv X_{i+3}$ with
$i=1,2,3$. This constraint is preserved by the equations of motion
and therefore if it is initially obeyed (as is the case in what
follows) it will be obeyed at all times. The equations of motion
are \bea
  \label{eq:motion1}
  \ddot{X}_i &=& \left\{ X_j,\left\{X_j, X_i \right\}\right\}+
   \left\{ Y_j,\left\{Y_j, X_i \right\}\right\}\nn \\
  \ddot{Y}_i &=& \left\{ X_j,\left\{X_j, Y_i \right\}\right\}+
   \left\{ Y_j,\left\{Y_j, Y_i \right\}\right\}
\eea
 We now use the ansatz of a rotating spherical membrane in analogy
 with the matrix membrane ansatz  given in \cite{HS}:
\bea\label{anz} X_i &=& r_i (t) e_i(\theta, \phi)\nn \\ Y_i &=&
s_i (t) e_i(\theta, \phi) \eea where the generators
$e_i(\theta,\phi)$ are defined as \bea e_1&=&\sin\theta \;
\sin\phi \nn \\ e_2&=&\sin\theta \; \sin\phi \\ e_3&=&\cos\theta
\nn \eea
 satisfy
the relations \be \left\{ e_i,e_j \right\}=-\epsilon_{i,j,k} e_k
\ee Using now the ansatz (\ref{anz}) in the equations of motion
(\ref{eq:motion1}) we obtain the differential equations obeyed by
the functions $r(t)$, $s(t)$
\begin{eqnarray}
  \label{eq:s-r}
  \ddot{r}_i&=&-(r^2+s^2-r_i^2-s_i^2)r_i \\
  \ddot{s}_i&=&-(r^2+s^2-r_i^2-s_i^2)s_i
\end{eqnarray}
where $r^2=r_1^2 +r_2^2 + r_3^2$, $s^2=s_1^2+s_2^2+s_3^2$. The
solution of (\ref{eq:s-r}) is of the form
 \begin{eqnarray}
  \label{eq:s-r-sol}
r_i &=& R_i \cos(\omega_i t + \phi_i) \\ s_i &=& R_i \sin(\omega_i
t + \phi_i)
\end{eqnarray}
with
\be
\omega_i^2 = R^2 - R_i^2 \ee where $R^2 = R_1^2 + R_2^2 + R_3^2$.

We observe that all the relations we obtained for the ansatz 
(\ref{anz}) are identical with those of ref.\cite{HS}  for the 
matrix model solution of a bound state of $N D_{0}/D_{2}$-branes 
where the three functions $ e _{i}(\theta,\phi)$ are replaced by 
N-dimensional representational matrices $ 
J_{\imath}(\imath=1,2,3)$ of $SU(2)$. This unique isomorphism is 
due to the existence of an $SU(2)$ subgroup of the infinite 
dimensional area preserving group of the sphere ($sDiff(S^2)$). It 
is known that there is no other finite dimensional subalgebra of 
$(sDiff(S^{2})$. As we shall see the stability analysis of the 
spherical membrane solution follows an isomorphic pattern with the 
matrix model solution . We point out that in ref\cite{HS} the 
matrix solution was found to be stable under a restricted set of 
the $l=1$ pertutbations. In the following we extend their analysis 
for every value of $l$ and we complete also the case $l=1$. The 
variational equations that correspond to the splitting in eq. 
$(4.3)$ between $X_{i}$ and $Y_{i}$  are: \bea \label{var} { 
\ddot\delta X_{i}} \; &=& \; \left\{ \delta X_{j}, \left\{X_{j}, 
X_{i}\right\}\right\} + \left\{ X_j ,\left\{\delta X_j, 
X_i\right\}\right\} \nn \\  &+& \left\{ X_j, \left\{ X_j, \delta 
X_i \right\}\right\} + \left\{ \delta 
Y_j,\left\{Y_j,X_i\right\}\right\}\nn \\ &+& 
\left\{Y_j,\left\{\delta Y_j,X_i \right\}\right\} + \left\{ 
Y_j,\left\{ Y_j,\delta X_i\right\}\right\}\eea The corresponding 
perturbation for $\delta Y_i$s and $ Y_i$s satisfy equations that 
are obtained by exchanging  $ \delta X_i \leftrightarrow \delta 
Y_i , X_i \leftrightarrow Y_i$ in eq(\ref{var}). The equations of 
motion imply the validity of the constraint at all times \beq 
\label{constr}\{ \dot{X}_i, X_i\} + \{ \dot{Y}_i , Y_i \} =0 \eeq 
This is obtained by taking the time derivative of 
eq.(\ref{constr}) and by applying the equations of motion and the 
Jacobi identity. By expanding a configuration which at $t=0$ is 
consistent with the constraint (\ref{constr}) around any classical 
solution we see (by using only the linearized eqs.(\ref{var})) 
that the variation $\delta X_i$ and $\delta Y_i$ satisfy the 
constraint \beq \{ \delta \dot{X}_i, X_i\} + \{ \dot{X}_i ,\delta 
X_i \} +\{{\delta {\dot Y}}_i, Y_i\}+ \{{\dot Y}_i, \delta Y_i\} = 
0 \eeq for all times. In order to proceed with the study of these 
variational equations we observe that: \beq \label{lambda} \{ e_i, 
Y_{lm} \} = i \hat{L}_i Y_{lm} \eeq where $\hat{L}_i$ are the 
angular momentum operators of Quantum Mechanics in spherical 
coordinates. In the N-dim. representation ($N=2l+1$) the right 
hand side is given by : \beq \label{Mi} \hat{L}_i Y_{lm}\;=\; 
\sum_{m^{'}} ( L_i)_{mm^{'}} Y_{lm^{'}} \eeq where 
$(L_i)_{mm^{'}}$ are the matrix representations of $SU(2)$. In 
what follows on the basis of the previous argument it is enough to 
consider the specific variation \beq \label{Ni} 
 \delta
X_{i}(t) = \sum_{m} \epsilon ^{m}_{i}(t) Y_{lm},\;  \delta 
Y_{i}(t)= \sum_{m} \zeta^{m}_{i} (t) Y_{lm}\eeq with initial 
conditions $\epsilon_{i}(0)=0 ,\zeta_{i}(0)=0$ but with 
$\dot{\epsilon}_{i}(0)\neq 0 ,\;  \dot{\zeta}_{i}(0) \neq 0$. As a 
result the constraint equation is satisfied at $t=0$. 

 In order to study the stability of this solution we
consider the following general form of perturbations \bea
\label{pertanz} \delta X_i (t)&=&\sum_{l,m} \epsilon_i^{lm}(t)
Y_{lm}(\theta,\phi) \nn \\ \delta Y_i (t)&=&\sum_{l,m}
\zeta_i^{lm}(t) Y_{lm}(\theta,\phi) \eea We now use the fact that
\be \left\{ e_i,Y_{lm}(\theta, \phi) \right\}=i {\hat L}_i
Y_{lm}(\theta, \phi) \ee where ${\hat L}_i$ is the angular
momentum differential operator. This  implies that \beq  \left\{
e_i,Y_{lm}(\theta, \phi) \right\} = \sum_{m'} 
a_{lm'}Y_{lm'}(\theta, \phi) = i \sum_{m'} (L_i)_{mm^{'}} 
Y_{lm^{'}}\eeq where $L_{\imath}$ are the angular momenta in the 
representation $l=(N-1)/2$. A crucial observation is that the sum 
involves spherical harmonics of the same $l$ as the spherical 
harmonic in the Poisson bracket. This decouples the various $l$ 
fluctuation modes and simplifies the differential equations obeyed 
by the modes $\epsilon_i^{lm}$ and $\zeta_i^{lm}$.  This feature 
is specific to the particular background solution of the spherical 
membrane. 

The equations obeyed by the fluctuation modes $\epsilon$ and 
$\zeta$ may be written as \bea \label{ezeq} {\ddot \epsilon}_i + 
R^2 l(l+1)\epsilon_i &=& R^2 \cos\omega t T_{ij}[\epsilon_j \; 
cos\omega t + \zeta_j \; sin\omega t]\nn \\ 
 {\ddot \zeta}_i + R^2 
l(l+1)\zeta_i &=& R^2 \sin\omega t T_{ij}[\epsilon_j \; cos\omega 
t + \zeta_j \; sin\omega t] \eea  where \be \label{tdef} 
T_{ij}=L_i L_j -2 i \epsilon_{ijk} L_k \ee and $L_i$ is the 
angular momentum operator. 

We now perform a rotation and define the new variables $\theta_i$ 
and $\eta_i$ \bea \theta_i &\equiv & \epsilon_i \cos\omega t + 
\zeta_i \sin\omega t \\ \eta_i &\equiv & -\epsilon_i \sin\omega t 
+ \zeta_i \cos\omega t \eea The equations obeyed by the new 
variables $\theta$ and $\eta$ may now be shown to be \bea {\ddot 
\theta}_i -2\omega {\dot \eta}_i + [R^2 l(l+1)- \omega^2]\theta_i 
- R^2 T_{ij}\theta_j &=& 0 \\ {\ddot \eta}_i -2\omega {\dot 
\theta}_i + [R^2 l(l+1)- \omega^2]\eta_i &=& 0 \eea where from the 
equation of motion of the background solution we have $\omega^2 = 
2 R^2$. Using this relation and defining the rescaled time 
variable $\tau=Rt$ we obtain the system \bea \label{theeq} {\ddot 
\theta} -2\sqrt{2} {\dot \eta} + [l(l+1)- 2]\theta & = & T\theta 
\nn 
\\ {\ddot \eta} -2\sqrt{2} {\dot \theta} + [l(l+1)- 2]\eta &=& 0 \eea 
where the time derivative is with respect to the new variable 
$\tau$ and we have suppressed indices. To investigate the 
stability we now use the ansatz
 \be
\left( 
\begin{array}{rl} \theta \\ \eta \end{array} \right)=e^{i \lambda \tau}
\left( 
\begin{array}{rl} a \\ b \end{array} \right)  \ee 
in the system (\ref{theeq}) to obtain the equations for $a$ and 
$b$ \be b={{2\sqrt{2} \lambda a}\over {i [l(l+1)-2-\lambda^2]}} 
\ee and \be \label{teigv} T a = [l(l+1) - 2 - \lambda^2 - 
{{8\lambda^2}\over {l(l+1) - 2 - \lambda^2}}]a \ee Therefore the 
problem of finding if $\lambda$ has an imaginary part (which would 
imply instability) has been reduced to solving the eigenvalue 
problem of the $3 (2l+1)\times 3(2l+1)$ Hermitian matrix $T$.  In 
order to solve this problem we will use the spectral theorem of 
algebra as follows: We expand the matrix $T$ into a complete set 
of three projector matrices and read from the coefficient of each 
term the eigenvalues which have degeneracy equal to the trace of 
each projector. The total number of eigenvalues should add up to 
$3 (2l+1)$ which is the dimensionality of $T$. Eigenverctors can 
also be found by acting with each one of the  projectors on any 
vector on the large space $3 (2l+1)$. The derivation of the 
explicit form of a {\it complete set} of eigenvectors however is a 
non-trivial task. 

Defining the two $3(2l+1)\times 3(2l+1)$ Hermitian matrices \bea 
P&=&{1\over {l (l+1)}} L_i L_j \nn \\ Q&=&i\epsilon_{ijk}L_k \eea
we observe that $P$ is a projector {\it ie} $P^2=P$ and $Q$ 
satisfies 
\be
Q^2 = l(l+1) (I-P) + Q \ee 

It is straightforward to show that $T$ may be expressed as: \be 
\label{texp} T=[l(l+1)-2]P+2lR_+ -2(l+1)R_- \ee where $P$, $R_+$ 
and $R_-$ are orthocanonical projectors with the usual properties 
$R_+ R_- =PR_+ = P R_- = 0$ and $R_+^2 = R_+$, $R_-^2 = R_-$. They 
are defined as \bea R_- &\equiv & {1\over (2l+1)}[(l+1)(I-P) - 
(I-Q)] \\R_+ &\equiv & {1\over (2l+1)}[l (I-P) + I-Q] \eea  From 
the spectral expansion of T (\ref{texp}) it becomes clear that its 
eigenvalues are $q_1=[l(l+1)-2]$, $q_2=2l$ and $q_3=-2(l+1)$ with 
multiplicities  given by the traces of the corresponding 
projectors ie $2l+1$, $2l+3$ and $2l-1$.  The corresponding 
eigenvectors are given by the set of $Pv$, $R_+ v$ and $R_- v$ 
where $v$ runs over ${\cal R}^{3 (2l+1)}$ 

Given now the eigenvalues of $T$ $q_i$ we are in position to use 
equation (\ref{teigv}) and find the form of the corresponding 
eigenfrequencies $\lambda_i$.  We must solve the algebraic 
equation \be l(l+1) - 2 - \lambda_i^2 - {{8\lambda_i^2}\over 
{l(l+1) - 2 - \lambda_i^2}}]=q_i \ee which leads to a pair of 
solutions for each $\lambda_i^2$. These solutions are \bea 
\lambda_{1a}^2 &=& 0, \;\;\;\;\;\; \lambda_{1b}^2=l^2 +  l + 6 \\ 
\lambda_{2a}^2 &=& l^2-3l+2, \;\;\;\; \lambda_{2b}^2=l^2 +  3 l + 
2 
\\ \lambda_{3a}^2 &=& l^2 - l, \;\;\;\; \lambda_{3b}^2=l^2 +  5 l + 6 
\eea It is obvious that all $\lambda_i^2$ are non-negative and 
therefore the eigenfrequencies $\lambda_i$ are all real. This 
implies that the membrane solution studied is stable to first 
order in perturbation theory. 

Perturbations of the classical solutions along the $7,8,9$ 
dimensions can be parametrized as \be \delta Z_i \equiv \delta 
X_{i+6} \ee with $i=1,2,3$ and \be  \ddot{\delta Z_i}_i= \left\{ 
X_j,\left\{X_j, \delta Z_i \right\}\right\} + \left\{ 
Y_j,\left\{Y_j, \delta Z_i \right\}\right\} \ee For the 
spherically symmetric membrane using the ansatz \be \delta Z_i = 
\rho_i^m (t) Y_l^m (\theta, \phi) \ee we find \be \ddot{\rho_i^m} 
+ R^2 l(l+1) \rho_i^m = 0 \ee that is stable harmonic motion. 

The above considered fluctuations are more general than the 
constraint (\ref{E3}) would allow. This however does not 
invalidate our analysis since we have shown that even those 
generalized fluctuations do not include an instability mode and 
therefore this will also be true for the physical fluctuations. 

These results however can not be valid to all orders. It is well 
known that the ground state of the system studied is a string or 
point configuration and therefore finite size fluctuations will 
eventually lead to a decay to the vacuum. This metastability may 
also be seen by considering higher orders in perturbation theory 
where non-linear effects start to show up. 

We conclude our work by stressing the analogy between the membrane
stability analysis presented above and the corresponding one for
the matrix model $ N D_{0}/D_{2}$ spherical solution of
ref\cite{HS}. As discussed before spherical matrix and membrane
solutions are isomorphic \cite{KT}. Moreover the linearized
problems for the fluctuations preserve the same isomorphism due to
the specific spin $1$ form of the solution. The matrix solution is
a bound state of N $D_{0}$ branes attached on a $D_{2}$ brane
whose stability properties is obtained using our continuous
membrane investigation. Indeed one only has to replace the
perturbations  $\delta X_i \;=\; \sum_{m} \epsilon_{i}^{m} Y_{lm}$
by the matrix fluctuations $ \hat{ \delta X}_i \;=\; \sum _{m}
\epsilon ^{m}_{i} \hat {Y}_{lm} $. Here $\hat{Y}_{lm}$ are the
$SU(2)$ tensor spherical harmonics defined in eq(\ref{E28}). One
can easily check that the 2l+1 dimensional vectors
$\epsilon_{i}(\zeta_{i})$ satisfy the same eqs.(\ref{ezeq}). On 
the other hand higher order perturbations differ by terms of order 
$1/N$ for every $N$.

At the linearized level small fluctuations of various multipole 
deformations described by $l=1,2,...$ do not destabilize the 
calssical solution. On the other hand we know that the potential 
term in the Hamiltonian of the supermembrane has as global minimum 
configurations tensionless strings and points. Therefore there are 
finite size deformations which can lead to instabilities of the 
rotating classical solution. In order to determine the modes of 
instability one has to study the potential of the moduli space of 
finite size deformations. 

Upon completion of our work we were informed by K. G. Savvidy of 
analogous results in the matrix model \cite{ssnew} 

%\begin{equation}
%  \label{eq:motion}
%  J^{0ij}= \left\{ X^i,X^j \right\}~~,
%\end{equation}

%\begin{equation}
%\label{WeakCond} \dot{X}^i \ll 1 , \spa [X^i,X^j] \ll l_s^2 , \spa
%\ddot{X}^i \ll l_s^{-1} , \spa [\dot{X}^i,X^j] \ll l_s~.
%\end{equation}

\section{Acknowledgements}
We are grateful to K.G.Savvidy for informing us about his work. We
also acknowledge G. Athanasiou, I. Kiritsis and T. Tomaras for 
useful discussions.

\bibliographystyle{prsty}

\bibliography{bibliog}

\begin{thebibliography}{10}
\bibitem{TWSD} P.Townsend, Phys.Lett. {\bf B350}, 184(1995);\\
E.Witten,Nucl.Phys.{\bf B443},85(1995);ibid{\bf B460},335(1995);\\ 
J.H.Schwarz, Phys.Lett. {\bf B360}, 13(1995);\\ P.Townsend, 
e-print hep-th/961212;\\ M.Duff, Int.J.Mod.Phys.{\bf A11}, 
5623(1996). 
\bibitem{BFSS}T.Banks, W.Fischler, S.H.Shenker, and L.Susskind,
Phys. Rev. {\bf D55}, 5112(1997), hep-th/9610043;\\ 
N.Ishibasi,H.Kawai,Y.Kitazawa,A.Tsuchiya,Nucl.Phys.{\bf B498},
467(1997), hep-th/9612115;\\ N.Taylor," Matrix Theory and M 
Theory" in NATO Lectures , hep-th/ 0001016. 
\bibitem{RCM} N.Taylor and M.Van Raansdom, "Multiple $D_{0}$-branes in Weakly
Curved Backrounds, hep-th/9904005;\\ R.C.Myers, "Dielectric
Branes" hep-th/9910053;\\ A.Nicolai and R.Helling, "
Supermembranes and Matrix Theory", hep-th/9809103.\\ B de Wit,
"Supermembranes and Supermatrix Models", Lectures in the 6th
Hellenic Summer School, Corfu , Greece 1998, hep-th/9901051;\\
ibid "Supermembranes in Curved Superspace \& Near Horizon
Geometries"Lectures at the 22nd John Hopkins School,
hep-th/9902149.
\bibitem{KT} D.Kabat and N.Taylor, "Spherical Membranes in Matrix
Theory", Advances in Theoretical \& Mathematical Physics,vol.2,
181 (1998), hep-th/9711078.
\bibitem{HS} T.Harmark and K.G. Savvidy, "Ramond-Ramond Field
Radiation from Rotating Ellipsoidal Membranes", hep-th/0002157;\\
K.G.Savvidy, "The Discrete Spectrum of the Rotating Brane",
hep-th/0004113.
\bibitem{DKPS} M.Douglas, D.Kabat, P.Pouliot and S.Shenker,
Nucl.Phys. {\bf B485}, 85(1997), hep-th/9608024.
\bibitem{WHN} J.Hoppe, Ph.D. Thesis MIT(1981), B. de Wit,
J.Hoppe and H.Nicolai, Nucl.Phys.{\bf B305}, 545(1988);\\ B. de 
Wit, M.Lusher and H. Nicolai, Nucl.Phys.{\bf B320}, 135(1989) 
\bibitem{BST} E.Bergshoeff, E.Sezgin, P.K.Townsend, Phys.Lett.
{\bf B189}, 75(1987);\\ E.Bergshoeff, E.Sezgin, Y.Tanii,
P.K.Townsend, Ann.Phys. {\bf 199}, 340(1990).
\bibitem{F} E.G.Floratos, Phys.Lett.{\bf B228}, 335(1989);\\
D.B.Fairlie, P.Fletcher, C.K.Zachos, Phys.Lett.{\bf B218}
203(1989).
\bibitem{FIT} E.G.Floratos, J.Iliopoulos and G.Tiktopoulos,
Phys.Lett {\bf B217}, 285(1989).
\bibitem{N} H.Nicolai and R.Helling, " Supermembrane and matrix
Theory ", Lectures at the Trieste School on Non-Perturbative
Aspects of String Theory and SUSY Gauge Theories, March 1998,
hep-th/9809103.
\bibitem{FL} E.G.Floratos and G.K.Leontaris, Phys.Lett.{\bf
B464},30 (1999).
\bibitem{PT} P.Townsend , Phys. Lett. {\bf B373}, 68(1996);
ibid," Four Lectures on M-Theory", Trieste 1996, hep-th/961212.
\bibitem{Sch} J.Schwinger, "On Angular Momentum", U.S. Atomic
Energy Commission, NYO-3071, 1952.
\bibitem{W96} E.Witten, Nucl.Phys.{\bf B460} (1996)335,
hep-th/9510135.
\bibitem{Ham}M.Hamemermesh, "Group Theory and its Application to
Physical Problems", Dover Publs. 1962.
\bibitem{Ba} G.Baym, "Lectures on Quantum Mechanics" Benjamin
1974.
\bibitem{Ar} V. Arnold, "Equations Differentielles Ordinaire", Ch.3, p.200 Ed.MIR-MOSCOU 1974. 
\bibitem{ssnew} K. G. Savvidy and G. K. Savvidy, 
%``Stability of the rotating ellipsoidal D0-brane system,''
hep-th/0009029. 

\end{thebibliography}

\end{document}